\documentclass[aps, reprint,groupedaddress,longbibliography,superscriptaddress]{revtex4-2}
\usepackage{graphicx, tabularx, longtable, nicefrac, amsmath,enumerate, supertabular, eurosym}
\usepackage{hyperref,xcolor}
\hypersetup{colorlinks = true, citecolor = blue!30!black, linkcolor = blue!30!black}
\usepackage{booktabs}
\usepackage{longtable}

\newcommand{\env}[1]{\texttt{#1}}
\begin{document}

\title{How do physics students evaluate artificial intelligence responses on comprehension questions? A study on the perceived scientific accuracy and linguistic quality}%

\author{Merten Nikolay Dahlkemper}
\email{merten.dahlkemper@uni-goettingen.de}
\affiliation{Faculty of Physics, Physics Education Research, University of Göttingen, Friedrich-Hund-Platz 1, 37077 Göttingen, Germany}
\affiliation{European Organization for Nuclear Research (CERN), Esplanade des Particules 1, CH-1211 Geneva 23, Switzerland}

\author{Simon Zacharias Lahme}
\affiliation{Faculty of Physics, Physics Education Research, University of Göttingen, Friedrich-Hund-Platz 1, 37077 Göttingen, Germany}

\author{Pascal Klein}
\affiliation{Faculty of Physics, Physics Education Research, University of Göttingen, Friedrich-Hund-Platz 1, 37077 Göttingen, Germany}

\date{\today}

\begin{abstract}
This study aimed at evaluating how students perceive the linguistic quality and scientific accuracy of ChatGPT responses to physics comprehension questions. A total of 102 first- and second-year physics students were confronted with three questions of progressing difficulty from introductory mechanics (rolling motion, waves, and fluid dynamics). Each question was presented with four different responses. All responses were attributed to ChatGPT, but in reality one sample solution was created by the researchers. All ChatGPT responses obtained in this study were wrong, imprecise, incomplete, or misleading. We found little differences in the perceived linguistic quality between ChatGPT responses and the sample solution. However, the students rated the overall scientific accuracy of the responses significantly differently, with the sample solution being rated best for the questions of low and medium difficulty. The discrepancy between the sample solution and the ChatGPT responses increased with the level of self-assessed knowledge of the question content. For the question of highest difficulty (fluid dynamics) that was unknown to most students, a ChatGPT response was rated just as good as the sample solution. Thus, this study provides data on the students' perception of ChatGPT responses and the factors influencing their perception. The results highlight the need for careful evaluation of ChatGPT responses both by instructors and students, particularly regarding scientific accuracy. Therefore, future research could explore the potential of similar "spot the bot"-activities in physics education to foster students' critical thinking skills.

\end{abstract}

\maketitle

\section{Introduction}
On November 30, 2022, the Artificial Intelligence (AI) company OpenAI released the large-language model \textit{ChatGPT} \footnote{\url{https://chat.openai.com/}, accessed on 21/02/2023} to the public \cite{openai_chatgpt_2022}. Even though the use of generative AI in education was widely discussed within the community before \cite{kasneci_chatgpt_2023} since the release of ChatGPT, the discussion on how generative AI will change the education sector has gained public attention. The new quality of ChatGPT is that it is easily accessible for everyone, can be used in a wide field of applications, and the answers are often of an unmatched quality \cite{kortemeyer_could_2023,choi_chatgpt_2023,gilson_how_2022,west_ai_2023,gregorcic_chatgpt_2023}, even for highly specialized fields \cite{holmes_evaluating_2023}. 

ChatGPT and similar chatbots, which are one class of generative AI,  will be used by different stakeholders in education in various use cases, e.g., for generating practice problems, step-by-step solutions, summaries, and outlines of texts, to name only a few \cite{kasneci_chatgpt_2023}. Since these chatbots are designed not to have a particular use case, but to be a versatile tool, they are used by students for getting answers to all kinds of problems, among them answering factual and conceptual questions about physics \cite{kortemeyer_could_2023,west_ai_2023}.

However, as long as chatbots are solely based on large language models without access to a structural knowledge base, the generated responses are based on a plethora of different texts with which the algorithm was pre-trained. Such texts may include misconceptions, and faulty logic, but also partly correct explanations. Briefly said, they are a mixture of all kinds of texts on that topic present in the training data, which --- in the case of ChatGPT --- includes parts of the world wide web up to a certain date (e.g., the year 2021 for the current version of ChatGPT). And even though future models will use results from web searches to base their responses, they might still be faulty and exhibit similar flaws as described above. 

So, when physics students use ChatGPT or a similar chatbot to solve physics questions, they are confronted with different challenges. First, the model may not always provide accurate or complete responses (see above). While ChatGPT has access to a vast amount of information, its responses are not perfect and may not exhibit the same level of understanding or depth of knowledge as a human expert in the field of physics. Furthermore, the responses generated by ChatGPT may be based on prompts that exhibit a wrong understanding of the context or nuances of the particular physics problem, which could lead to incorrect or irrelevant responses. Students need to provide detailed information and context about the problem they are trying to solve to receive a useful response from the model \cite{gregorcic_chatgpt_2023}. Finally, the responses might not be factually wrong but do not explain anything. Given a well-written text from a supposedly authoritative source, this might create an illusion of understanding where students believe they have grasped the concept, but in reality, they only memorize a response without a full comprehension of the problem and its solution. This can hinder their ability to learn and apply the concepts in future problem-solving situations.

Thus, meaningful and fruitful use of generative AI tools like ChatGPT by students requires reflection and digital literacy, e.g., to identify whether ChatGPT responses are correct and helpful for answering the question. In the present study, we investigate to which extent students adequately perceive the (a) scientific accuracy and (b) linguistic quality of ChatGPT responses to introductory physics questions. By doing so, the study contributes to understanding student interactions with generative AI in education and how these interactions impact students' learning experiences. Connected to this line of research are plenty of new research questions, including the impact of using generative AI on students' motivation, engagement, and attitudes toward learning, and the potential benefits and limitations of integrating generative AI into physics curricula and pedagogical practices.

\section{State of research and research questions}
 
\subsection{Chatbots and their use in education}

Chatbots have been around for several decades. The first chatbot that was introduced to the public was ELIZA \cite{weizenbaum_eliza_1966} already in 1966. This chatbot used a set of rules on how to react to specific keywords within the prompts to emulate a psychotherapist's answers to questions. Today's generative AI tools work on an entirely different model. They use enormously large amounts of training data to build a statistical model of language. 

Chatbots have various applications for education. According to a recent review \cite{okonkwo_chatbots_2021} based on 53 studies from 2016 to 2021, the most common application that was investigated in the literature was teaching and learning, i.e., for instructors to deliver course content, or to provide students with engaged experience, and for students to ask questions, and to get individualized help. Besides teaching and learning, chatbots can assist students in research and development, advise students (e.g., for decisions about career and education), as well as assist instructors with assessment and administrative tasks. 

Besides possible applications, also the acceptance of using AI tools, such as chatbots, amongst instructors and students has been investigated \cite{chocarro_teachers_2023, chatterjee_adoption_2020}. It was found that the perceived risk of the technology has a negative impact on the attitude toward the adoption of AI while the effort expectancy, i.e., the perceived ease of use has a positive impact. The attitude towards adopting AI technology acts as a strong predictor variable for the behavioral intention of using AI technology in education \cite{chatterjee_adoption_2020}. 

The recently published chatbot ChatGPT is a refined version of the large-language models \textit{InstructGPT} \cite{ouyang_training_2022} and \textit{GPT-3.5}. These models are based on a large body of internet data and were trained with different methods of human feedback \cite{openai_chatgpt_2022}. Since its release on November 30th, 2022, it has been updated several times to react to user feedback \cite{openai_chatgpt_2022-1}. However, as the developers of ChatGPT write on their website \cite{openai_chatgpt_2022}, "ChatGPT sometimes writes plausible-sounding but incorrect or nonsensical answers. Fixing this issue is challenging [...]". After conducting this study and while writing this paper, a new model, \textit{GPT-4} \cite{openai_gpt-4_2023}, was introduced by OpenAI, which is not only capable of processing language but also multimodal input. Also, the performance of GPT-4 when responding to physics problems is reportedly much better than that of GPT-3.5 \cite{west_advances_2023, holmes_evaluating_2023}. 

The unmatched quality of ChatGPT responses is the reason why it has reached much media attention \cite{lock_what_2022,roose_brilliance_2022,verma_what_2022} and is currently an arising object of interest for research across many fields. Several researchers have examined the quality of ChatGPT responses to exam questions, notably for medical \cite{gilson_how_2022}, pharmaceutical \cite{fergus_evaluating_2023}, law \cite{choi_chatgpt_2023}, economics \cite{geerling_chatgpt_2023}, and physics exams \cite{kortemeyer_could_2023,west_ai_2023}. They found that ChatGPT can answer enough questions sufficiently correctly to pass standardized exams in various fields. However, a recent review of the performance of ChatGPT on multiple choice-based exams concluded that, overall, ChatGPT performs only modestly on these tasks, especially when they included problem-solving, transfer tasks, or maths problems \cite{newton_chatgpt_2023}. It was concluded that ChatGPT can outperform humans especially on mere recall questions, but not on questions that ask for problem-solving or transfer \cite{newton_chatgpt_2023, geerling_chatgpt_2023}. Especially for physics questions, the studies noted that ChatGPT exhibits errors that resemble those of novice physics learners while not showing any signs of metacognition, i.e., ChatGPT presents potentially false information as if it was a fact \cite{kortemeyer_could_2023,west_ai_2023,gregorcic_chatgpt_2023}. This issue has been reportedly partly resolved in the latest version of ChatGPT (GPT-4), while also this model still produces erroneous responses in certain cases \cite{west_advances_2023}. In a recent pilot-study \cite{bitzenbauer_chatgpt_2023}, two activities with ChatGPT regarding the critical thinking skills of students were implemented in the physics classroom. The study gave hints towards a positive influence of this activity towards ChatGPT. 

\subsection{Illusion of understanding}
\label{sec:illusion}
Since misconceptions might be included in responses from ChatGPT (cf. \cite{kortemeyer_could_2023,west_ai_2023}), it is crucial to have an understanding of the topics concerned when working with those responses. However, the self-assessed understanding often does not match with an actual understanding \cite{prinz_double_2018}. In a recent study \cite{kulgemeyer_misconceptions_2023}, it was shown that students who watched explanatory videos containing misconceptions developed an equal belief of understanding as students who watched similar videos without these misconceptions.
This research can be applied to the use of chatbots which might give faulty explanations to physics questions. If students use these chatbots they might believe in faulty explanations because they might sound plausible, are perhaps written engagingly, or the design of the webpage could add to the illusion of credibility.

\subsection{Research questions}
\label{sec:rq}
As far as we know, besides testing the chatbot ChatGPT in the context of physics tasks, no survey has been conducted yet that faces the students' perception of ChatGPT responses. However, as ChatGPT is an emerging tool also in the educational context, one needs to consider that (physics) students use and will use ChatGPT for solving physics questions. Building on the partly faultiness of ChatGPT responses, it would be important that students critically evaluate the responses provided by ChatGPT to identify errors and incompleteness of responses. Therefore, our research interest is whether physics students can adequately evaluate ChatGPT responses to physics questions. Accordingly, our main research question is:\\
\textbf{1. How do introductory physics students rate ChatGPT responses to phenomenological mechanics comprehension questions in terms of \textit{scientific accuracy} and \textit{linguistic quality} compared to a (masked) sample solution?}\\

As written above (section~\ref{sec:illusion}), the illusion of understanding refers to the tendency of students to overestimate their understanding of a topic, particularly when they have limited knowledge or incomplete information. This could result in judgment errors. In the context of rating ChatGPT responses, students who have limited content-related expertise might be more likely to give higher ratings to responses that they believe are correct, even if they do not fully understand the topic. This could lead to inaccurate evaluations of the quality of ChatGPT responses. Therefore, it is important to examine how the students' self-estimated content-related expertise influences their rating of ChatGPT responses, as this could help to identify potential sources of bias in the evaluation process and shed light on the accuracy of the ratings. Therefore, the second research question acknowledges that the evaluation of ChatGPT responses might depend on the prior knowledge regarding the physics question itself: \\
\textbf{2. What impact does the students' self-estimated content-related expertise have on this rating?} \\

The linguistic quality of a response might also influence how accurately it is rated. A response that is well-written and clearly explains the scientific concepts might be more likely to be rated as accurate, even if it contains errors or misconceptions. Therefore, this research question about the interdependence of scientific accuracy and linguistic quality is important to investigate because it can help to identify the impact of linguistic quality on the illusion of understanding. 
The third research question, therefore, investigates the interdependence of scientific accuracy and linguistic quality.\\
\textbf{3. Do any (potential) discrepancies regarding the rating of scientific accuracy still exist when the rated linguistic quality of the responses is considered and corrected for?}

\section{Methods and materials}

The instrument consisted of two parts. In the first part, demographic information as well as information on the students' attitudes toward AI were collected. In the second part, the students were presented with three physics questions, and subsequently, for each question three responses were provided by ChatGPT as well as one sample solution provided by the authors that was also labeled as a ChatGPT response. The students should then assess the perceived quality of the responses as a solution to the question. The instrument was given to the students in German language (see Table \ref{tab:instrument-translation} in the appendix for the original and translated form of the questions). In this section, it is described in the translated form.

\subsection{Survey instrument}
\label{sec:survey}
The questionnaire was implemented as an online survey in the open-source tool LimeSurvey. After a front page on which the students were briefly informed about the goal of the survey (perception of ChatGPT responses to physics questions among physics students) and were asked for their consent to data collection.
In the first step, the students were asked about their subject of study (\textit{physics major, physics teacher training, other}), their current year of study, and what gender they identify as. They were also asked to rate three statements about their expectation of the performance of AI (\textit{A chatbot allows me to get accurate answers. Answers from a chatbot are not always correct.}) as well as their study-related attitude towards AI in general (\textit{Artificial intelligence (AI) is useful for my studies.}) These items were adapted from a survey instrument on the Unified Theory of Acceptance and Use of Technology \cite{chatterjee_adoption_2020}. The items are answered on a 5-point Likert scale (\textit{strongly disagree, disagree, neither agree nor disagree, agree, strongly agree}), but the students could also skip the question (\textit{don't know/prefer not to say}).

Subsequently, they were asked whether they have heard about the chatbot ChatGPT (\textit{Have you ever heard of ChatGPT?}) and if so, how often (\textit{never, once, 2-5 times, more than 5 times}) they have used it so far to work on physics exercises in their studies and for other (e.g., private) use cases respectively. After that, they were given a brief explanation of what ChatGPT is and what they will have to do in the remainder of the survey. We only explained ChatGPT at this stage to not bias the students' responses to the previous questions.

In the second step, students were presented with three physics comprehension questions one after another (further described in section~\ref{TasksAnswers}) and were asked to give an estimation of their level of performance when the questions would have been part of an exam (\textit{Without solving the question yourself: Estimate how many points you would have received on this question in a written exam (0-6 points)}). This form of self-assessment comes close to what \cite{colliver_self-assessment_2005} has described as the “guess your grade” model of self-assessment research. Even though this form of self-assessment usually has clear disadvantages (e.g. systematic overestimation of one's abilities), we use the students' ratings as an estimation of their potential performance on these questions. We did not want to let the students solve the questions themselves, firstly because of time constraints and secondly because their opinion about the presented responses would have been influenced too much. However, it is likely that students started to craft a solution in their minds. In follow-up studies, one could use other indicators of performance that are not based on self-assessments. 

In the third step, after each question and the respective self-assessment, the students were presented successively with four responses to these physics comprehension questions, all of them labeled as written by ChatGPT. However, only three responses were provided by ChatGPT, the fourth response was a sample solution written by the authors and containing a correct and complete explanation of the question. The students were asked to rate each response regarding five criteria on a 5-point Likert scale (\textit{very low, low, medium, high, very high}). The criteria address both the scientific accuracy (\textit{factual correctness} and \textit{degree of completeness}) and the linguistic quality (\textit{comprehensibility} and \textit{quality of the language}) of the responses. Additionally, the students were asked whether the presented response would be an appropriate sample solution for the question (\textit{suitability as a sample solution}). As will be described in section~\ref{sec:red}, the first two and the last item would later be combined into the scale \textit{scientific accuracy} whether the third and fourth item would be combined into the scale \textit{linguistic quality}. The students also had the opportunity to take another look at the question itself. All four responses to be rated for each question were displayed to the participants in a randomized order to minimize the effect of judgment errors (e.g., that the participants rate the first displayed response better or worse than later responses as they have no comparison in the first place). The same procedure was applied to the other two physics questions.

The study variables \textit{scientific accuracy} and \textit{linguistic quality} were collected by rating scales, so that strictly speaking one would have to refer to the "perceived" \textit{scientific accuracy} and "perceived" \textit{linguistic quality}. 
Also, to keep the test load limited, we only used five items to assess these constructs, since the survey should be feasible within a 30-minute time frame at the end of a lecture to gather answers from as many students as possible. With 12 assessed responses, this would not be possible with a more detailed scale to answer for each response.

The survey ended with an open text field in which the participants could write any comment regarding the survey if necessary (\textit{Would you like to tell us something concerning the survey in conclusion? Then you can do so here...}).

\subsection{Physics questions and responses} \label{TasksAnswers}

We formulated three physics comprehension questions in German language which belong to three different topics of the lecture on mechanics for first-year physics students. These topics were rolling motion, waves, and fluid mechanics. The questions were constructed such that they belong to three progressing levels of difficulty. The rolling motion question (question 1) was a mere repetition for most of the students as they have seen the same question (and its solution) a few weeks before as an exercise question. The wave question (question 2) was an application of knowledge the students have learned a few weeks before. And the fluid mechanics question (question 3) required a transfer of knowledge to an unknown and more complex problem which required students to combine several concepts they have learned. 

For each question, five different responses were generated using the Jan 9 release of ChatGPT \cite{openai_chatgpt_2022-1}. For each response, a new chat was opened, so the program was not influenced by its previous responses. The program was prompted with the German questions formulated before and the responses were taken as is, with two exceptions: Prompted with the first question, the program responded in English on one occasion. In this case, the program was subsequently asked to write that response in German. The response to that question was then taken as the final response. Prompted with the second question, the responses of the program were significantly longer than the responses to the other two questions. Therefore, the program was asked to shorten its response to one paragraph. Additionally, one correct and complete sample solution per question was formulated by the researchers (referred to as sample solution). The sample solution was given in the same style as the ChatGPT responses, i.e., as pure, explanatory text. For each question, three ChatGPT responses were selected to account for the variety of different responses given by the program and at the same time to keep the rating effort for the students acceptable. The selection was made during the pilot phase in a process described in section~\ref{sec:validation}. All questions, ChatGPT responses, and sample solutions as well as their English translations are provided in Table~\ref{tab:qandr} in the appendix. Exemplary, question 3 including the three selected ChatGPT responses and the sample solution in a translated form can be found in Table~\ref{Task3}.

\begin{table*}[htb]
\caption{Question and responses for question 3 (fluid mechanics) with a high level of difficulty. Responses 3.3 to 3.5 were generated by ChatGPT in German and translated afterward, response 3.6 is the sample solution created by the authors.}
\begin{ruledtabular}
\begin{tabular}{p{.09\textwidth}p{.85\textwidth}}
Question 3&\hangindent=.3cm Paper scraps are lying on the bottom of a glass filled with water. You stir the water with a spoon. After removing the spoon, you observe that the paper scraps rise and collect in the middle of the bottom after some time. Explain this phenomenon!\\\hline
R3.3&\hangindent=.3cm This phenomenon can be explained by the process of sinking and rising of air bubbles in the water. When the water is stirred with a spoon, air bubbles are stirred into the water, which collect on the paper scraps and cause them to float upwards. After some time, the paper scraps sink again as the air bubbles escape from the paper scraps and the paper scraps have their specific gravity again, which drives them back to the bottom. Therefore, the paper scraps collect in the middle of the bottom because the air bubbles are no longer present and the paper scraps have their specific weight.\\
R3.4&\hangindent=.3cm This phenomenon is explained by the so-called Bouyancy effect. When you stir the water with the spoon, a current is created that pushes the paper shreds upwards. However, the paper scraps sink again because their specific weight is greater than that of the water and they, therefore, sink back down. The paper scraps finally collect in the middle of the bottom, as they come to rest there due to the natural balancing of the currents.\\
R3.5&\hangindent=.3cm This phenomenon occurs because the paper initially has a higher density than the surrounding water, which causes it to sink. However, when it absorbs water and thus reduces its density, it rises. Eventually, however, it becomes heavier again and, due to its greater density, sinks to the bottom, where it gathers in the middle.\\
R3.6&\hangindent=.3cm The spoon causes the water to rotate so that a centrifugal force acts on the water. This force is countered by a frictional force due to the friction of the water with the bottom of the glass. The resulting velocity gradient creates a secondary flow superimposed on the rotational motion. This flow runs from the center of the bottom of the glass upwards, in higher layers towards the edge, at the edge of the glass downwards, and at the glass bottom towards the center. The paper scraps are carried along by this flow. After removing the spoon, the flow decreases due to friction and the shreds finally collect in the center of the bottom.
\end{tabular}
\end{ruledtabular}
\label{Task3}
\end{table*}

\subsection{Validation and pilot of the survey}
\label{sec:validation}
The questions and all initially generated responses (five responses provided by ChatGPT and the sample solution) were validated with physics faculty staff (referred to as \textit{experts} in the following) and physics students of higher years of study. The pilot had three different goals:
    (1) Based on the responses of all pilot participants, the perceived variety of the quality in the five ChatGPT responses was investigated to select those three ChatGPT responses for the main survey whose quality was assessed most differently to represent the quality continuum of ChatGPT responses. The three chosen responses chosen should reflect the range of responses obtained. Additionally, it served as a manipulation check by testing if the sample solution performs best.\\
    (2) With the experts, we also checked if the sample solution needed any revision. In case the experts rated one sample solution to be unsuitable as such, they were subsequently asked to state clearly what the reason was.\\
    (3) With the students of higher semesters, the comprehensibility of the survey text elements as well as the necessary amount of time for participation in the survey were piloted. Thus, time stamps were taken and the students answered four open questions regarding the comprehensibility of the survey text elements and the criteria in particular.\\
In total, the responses were rated fully by seven and additionally partly by three experts for Question 1, fully by three and additionally partly by two experts for Question 2, and fully by three experts for Question 3. Furthermore, five students (physics major or physics teacher-student, 3$^{\text{rd}}$ to 5$^{\text{th}}$ year) rated the 18 responses to all three questions. The pilot phase revealed the following:

    (1) The results of the validating and pilot phase were used to decide which ChatGPT responses are used in the final instrument.
    Generally, we implemented the ChatGPT responses into the instrument which were rated \textit{worst} and \textit{best} on average throughout all criteria and all participants in the pilot study. To choose a third response, we took additional qualitative criteria into account, such as the differentiability between different responses. For example, if all but one responses share a certain type of explanation, we implemented also the one that used another type of explanation. The precise reason for every response that was implemented or not implemented is given in Table \ref{tab:qandr} in the appendix. \\
    (2) The experts rated the sample solution highest in all five categories, so they can be assumed to be suitable as such. One expert rated the sample solution of Question 1 as unsuitable. Based on the written feedback, response 1.4 (sample solution) was improved by adding a missing condition (cf. Table \ref{tab:qandr}). Also, one expert rated the sample solution of Question 2 as unsuitable but did not give any reason, so we could not take that judgment into account in the revision of this solution. One expert rated the sample solution of question 3 as unsuitable but stated in the comment that they rushed through and probably didn't read the question correctly. Additionally, one pilot participant reached out after the main study had already begun and pointed out that mentioning "centrifugal force" in the sample solution to question 3 is an imprecision when describing the problem in the fixed frame, even though the term is often used colloquially in text books and lectures. However, the sample solution remained a valid and complete explanation of the phenomenon (cf. Table~\ref{Task3}). Since the term is still widely used in German introductory text books, it is unlikely that students judged the response due to this term.  \\
    (3) The students perceived the survey as largely comprehensible. They had difficulties in distinguishing between the formerly used criteria \textit{Sprachliche Verständlichkeit} (English translation: \textit{linguistic comprehensibility}) and \textit{Sprachliche Präzision} (English translation: \textit{linguistic precision}) for the rating of the different responses, so they were substituted by the already mentioned criteria \textit{Verständlichkeit} (English translation: \textit{comprehensibility}) and \textit{Sprachliche Qualität} (English translation \textit{quality of the language}). The students' pilots lasted 23-53~minutes ($M=36$~min, $SD=11$~min), so by shortening the survey as intended (only three instead of five ChatGPT responses per question) the intended time frame of 20-30~minutes was reached.\\

\section{Data collection}
Data were collected with the original German version of all physics questions, ChatGPT responses, sample solutions, and additional questions in the survey in two different physics courses at the University of Göttingen in the last week of winter term 2022/2023 on February, 8$^{\text{th}}$ and 10$^{\text{th}}$ 2023. The first course was the introductory experimental physics lecture about mechanics and thermodynamics ("Experimentalphysik 1") for physics majors and physics teacher students in their first semester. The course was held by the authors themselves and addressed the topics related to the three physics questions in this survey. Thus, for this target group the description of progressing levels of difficulty described in section \ref{TasksAnswers} applies. The participants were briefly instructed that ChatGPT is a new language-based AI tool that can also be used to solve physics questions. After that, they solely participated in the online survey which took most students less than 20~minutes ($M=14~$min, $SD=5~$min). Afterward, the students were provided the sample solutions.
In total, 84 students of this subgroup participated in this survey. Three participants were excluded from the data set because they have not rated all responses for at least one question and one further participant was excluded due to an unrealistic fast processing time of less than four minutes.

The second course in which the students participated in this survey was a third-semester experimental physics course ("Experimentalphysik III") about wave optics and atom physics. The lecture was held by another lecturer, but all students should have already participated in the first-year mechanics and thermodynamics course before. However, the description of progressing levels of difficulty does not necessarily apply to this subgroup, since they might not have seen the first question in the past. The data collection was organized in the same way as for the other subgroup. In total, 15 students of this subgroup participated in this survey fully and 3 further students partly. One participant was excluded from the further data analysis as not all responses for at least one question were rated.

So, in total 94 full participants and 3 additional partial participants were considered in the further data analysis. A description of this group of participants regarding their field of study, semester, and gender can be found in TABLE \ref{Participants}. 
\begin{table}
\caption{Overview of the number and characteristics of the participants in the two university physics courses in which the survey was conducted.}
\begin{ruledtabular}
\begin{tabular}{p{.35\columnwidth}p{.305\columnwidth}p{.305\columnwidth}}
&Course 1 (Experimentalphysik I)&Course 2 (Experimentalphysik III)\\\hline
Participants (total)&80 fully + 4 partly&15 fully + 3 partly\\
Participants (cleaned)&79 fully + 1 partly&15 fully + 2 partly\\\hline
Field of study&&\\
- physics major&59&15\\
- physics teacher&16&\\
- other&5 (math)&2 (math \& applied data science)\\\hline
Semester&&\\
- $1^{\text{st}}$ semester&75&\\
- $3^{\text{rd}}$ semester&3&17\\
- $>3^{\text{rd}}$ semester&2&\\\hline
Gender&&\\
- male&61&10\\
- female&15&5\\
- divers&&1\\
- no specification&4&1
\end{tabular}
\end{ruledtabular}
\label{Participants}
\end{table}

\section{Results}

We first report the descriptive results of the first part of the survey about the students' familiarity with ChatGPT and their general attitude towards AI (\ref{sec:fam-att}) and the students' scores on their self-assessment for the three physics questions they were given (\ref{sec:self}). Afterward, we describe the data reduction of the student's answers to the assessment of scientific accuracy and linguistic quality of the responses (\ref{sec:red}). Subsequently, we describe the results of this analysis regarding the three research questions stated in section~\ref{sec:rq}. First, we report the perceived scientific accuracy and linguistic quality (\ref{sec:perc}), then the impact of the self-assessment score on this assessment (\ref{sec:gap}), and finally the impact of the perceived linguistic quality on the perceived scientific accuracy (\ref{sec:impact}).

\subsection{Students' familiarity with ChatGPT and attitude towards AI}
\label{sec:fam-att}

The majority of the students had already heard of ChatGPT (84\%) before the survey was conducted, but only about half of them (48\%) had used the chatbot. Only a minority reported using the chatbot frequently (8\%). We also asked whether ChatGPT had ever been used in the context of physics questions, which was denied by 74\% of the students.

In terms of attitudes regarding the role of AI in education, the majority of students (50 out of 85) agreed strongly that AI-generated answers can be prone to errors, with a mean score of 4.27 out of 5 (SD = 1.11). However, when asked about the usefulness of AI chatbots for physics studies, the level of agreement was moderate, with a mean score of 3.47 (SD = 1.06). Additionally, students had a moderate level of expectation for accurate answers from AI chatbots, with a mean score of 3.13 (SD = 0.75). 

\subsection{Students' self-assessment of performance in different physics topics}
\label{sec:self}
\begin{figure*}[t!]
    \centering
    \includegraphics[width=\textwidth]{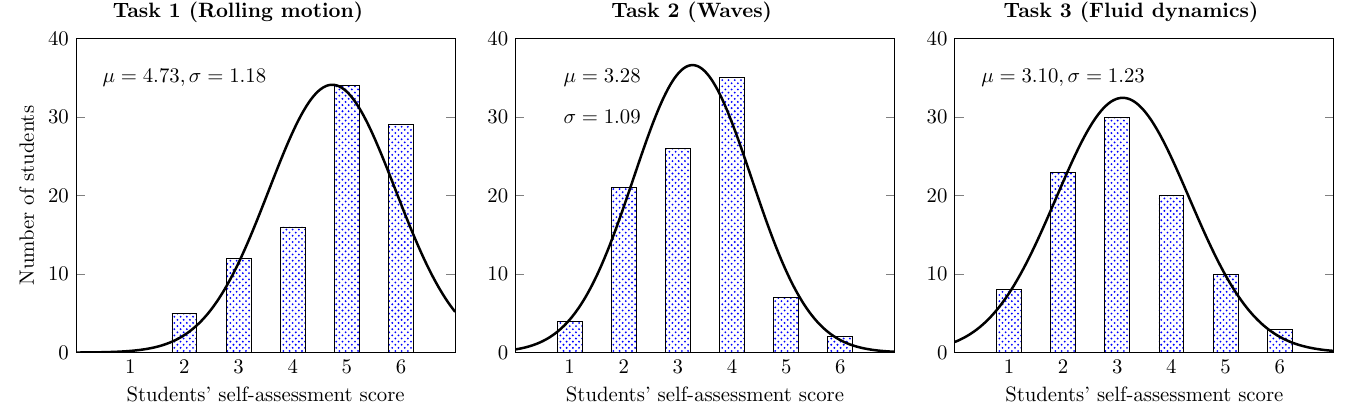}
    \caption{Distribution of students' self-assessment of their performance ("Without solving the question yourself: Estimate how many points you would have received on this question in a written exam (0-6 points)") regarding questions 1--3. }
    \label{fig:histos}
\end{figure*}

Students were required to evaluate not only the given responses to the physics questions based on various criteria but also estimate their performance on a scale from 0 to 6 points when they would have solved the questions under exam conditions. The histograms in Figure \ref{fig:histos} depict the self-assessments of the students for the three questions. They show that the students rated their knowledge higher for the first question (related to rolling motion; lowest level of difficulty) compared to the second question (waves; medium level of difficulty), and the third question (related to fluid dynamics; highest level of difficulty). So, the students' self-assessment is following our intended level of difficulty.

Since the intended level of difficulty might not necessarily hold for second-year students, we compared the self-assessment rating between the first- and second year students by using a t-test. For neither of the three questions, we found a significant difference in the students' self-assessment (Question 1: $t(94)=-.4,\ p = .72$; Question 2: $t(93) = .9,\ p=.40$; Question 3: $t(92)=.1,\ p=.89$).

\subsection{Data reduction and data preprocessing}
\label{sec:red}
The students evaluated each response based on five categories (i.e., factual correctness, degree of completeness, comprehensibility, quality of the language, and suitability as a sample solution).  Exploratory factor analysis revealed that the first two categories group together with the last one; hence, the ratings were averaged to represent the variable \textit{(perceived) scientific accuracy} for further analysis. This 3-item scale achieved a mean reliability of Cronbach's $\alpha= 0.80$ (see Table \ref{Reliabilitäten}). Further, the third and fourth item group together, representing the \textit{(perceived) linguistic quality} scale that obtains a mean Spearman-Brown coefficient of $r=0.65$. The exact values of the reliabilities can be found in Table \ref{Reliabilitäten}.

\begin{table}
\caption{Reliabilities for the two scales \textit{scientific accuracy} based on the three criteria \textit{factual correctness}, \textit{degree of completeness} and \textit{suitability as a sample solution} and \textit{linguistic quality} based on the two criteria \textit{comprehensibility} and \textit{quality of the language}. Cronbach's $\alpha$ (for $k=3$ items) and Spearman-Brown coefficient (for $k=2$ items; 94 to 96 degrees of freedom, depending on the response) are calculated.}
\begin{ruledtabular}
\begin{tabular}{p{.47\columnwidth}p{.245\columnwidth}p{.245\columnwidth}}
Question/Response&Scientific accuracy ($k=3$)&Linguistic quality ($k=2$)\\\hline
\textbf{Question 1 (rolling motion)}&&\\
R1.1 (ChatGPT)&.82&.61\\
R1.2 (ChatGPT)&.83&.63\\
R1.4 (sample solution)&.85&.78\\
R1.6 (ChatGPT)&.62&.66\\\hline
\textbf{Question 2 (waves)}&&\\
R2.1 (ChatGPT)&.73&.73\\
R2.2 (ChatGPT)&.86&.62\\
R2.4 (sample solution)&.89&.83\\
R2.6 (ChatGPT)&.81&.71\\\hline
\textbf{Question 3 (fluid dynamics)}&&\\
R3.3 (ChatGPT)&.78&.45\\
R3.4 (ChatGPT)&.84&.67\\
R3.5 (ChatGPT)&.70&.63\\
R3.6 (sample solution)&.86&.52\\\hline
Mean&\textbf{.80}&\textbf{.65}
\end{tabular}
\end{ruledtabular}
\label{Reliabilitäten}
\end{table}

\subsection{Analysis of perceived scientific accuracy and linguistic quality}
\label{sec:perc}
\begin{figure*}[t!]
    \centering
    \includegraphics[width=\textwidth]{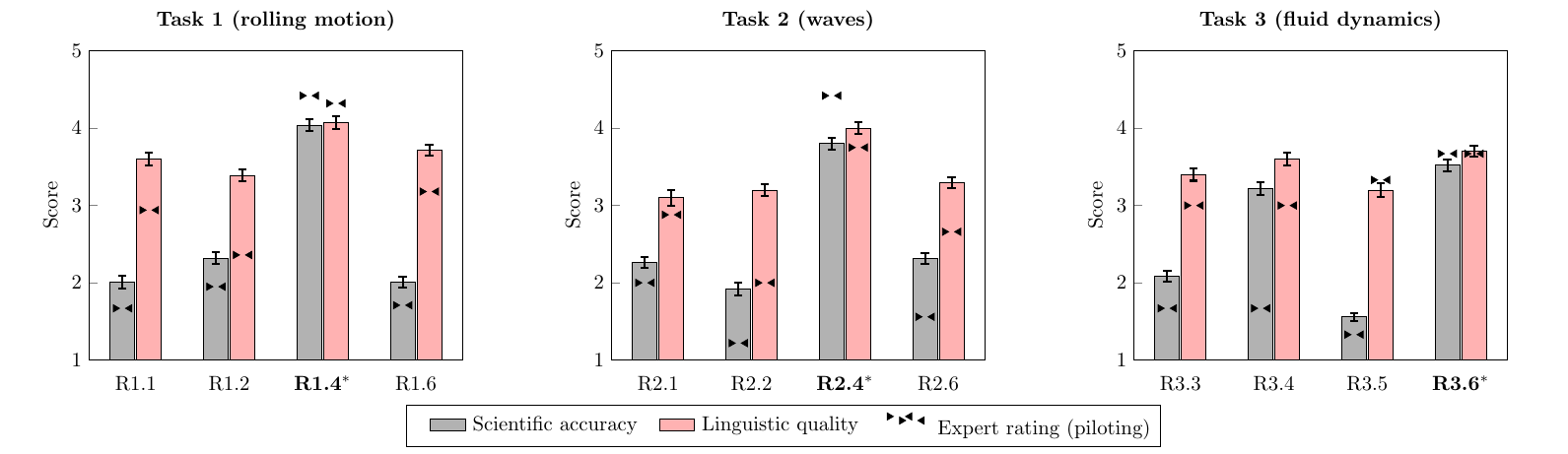}
    \caption{Students' perceived accuracy and linguistic quality of the four individual responses to each of the three questions. The correct sample solution is printed in bold (R1.4, R2.4, and R3.6). The experts' assessments from the pilot are also shown. The error bars indicate the standard error.}
    \label{fig:RM}
\end{figure*}

To answer the first research question, i.e., to evaluate the impact of the different responses presented and the assessment criterion (scientific accuracy vs linguistic quality) on the students’ ratings (see Fig. \ref{fig:RM}), a two-way repeated-measure ANOVA ($4\times 2$ ANOVA rm) was conducted for each of the three questions. 

For the first question, we found a statistically significant main effect of both factors, presented response [$F(3, 285)=99.9, p < 0.001, \eta^2_p=0.51$] and assessment criterion [$F(1, 95)=321, p < 0.001, \eta_p^2=0.77$], and also the interaction effect was significant [$F(3, 285)=114, p < 0.001, \eta^2_p=0.55$]. These results mean in particular that (1) the expert solution received the highest ratings among all responses ($M=4.03, SD = 0.81$ for scientific accuracy, $M=4.07, SD = 0.76$ for linguistic quality); (2) students rated the overall linguistic quality higher ($M = 3.69, SD=0.54$) than the scientific accuracy ($M=2.59, SD = 0.48$), and (3), there are consistently large differences in the ChatGPT responses concerning the assessment of the two criteria, whereas these differences are very small in the expert solution.

For the second question, similar results were obtained. We found a statistically significant main effect of both factors [$F(3, 279)=108, p < 0.001, \eta^2_p=0.54$ and $F(1, 93)=231, p < 0.001, \eta^2_p=0.71$ for the presented response and assessment criterion, respectively] and of the interaction [$F(3, 279)=44.6, p < 0.001, \eta^2_p=0.32$]. Again, the expert solution achieved higher ratings than the ChatGPT responses, the linguistic quality was overall assessed higher than the scientific accuracy, and the difference between both ratings is smaller for the expert solution compared to the ChatGPT responses. 

For the third question, the statistical analysis yielded similar results. The main effect of both factors is significant [$F(3, 276)=110, p < 0.001, \eta^2_p=0.55$ and $F(1, 92)=264, p < 0.001, \eta^2_p=0.74$ for the presented response and assessment criterion, respectively] as well as their interaction [$F(3, 276)=102, p < 0.001, \eta^2_p=0.53$]. Inspecting the descriptive data in Fig. \ref{fig:RM} (c) reveals that response R3.4 was rated closer to the expert solution than any other response in the data set. Post-hoc analyses show that the perception of linguistic quality does not differ between R3.4 and R3.6 [T(92)=1.09, p=0.28]; however, the expert solution has significantly higher ratings regarding the scientific accuracy than the ChatGPT response R3.4 [T(92)=2.82, p=0.01].

FIG. \ref{fig:RM} also shows that the experts (physics faculty) systematically rate both the scientific accuracy and the linguistic quality of the ChatGPT responses lower than the students, while at the same time, the sample solutions tend to receive higher ratings. In the next section, we analyze the gaps between the ratings of the sample solution and the chatbot responses in more detail. 

\subsection{Analysis of the gap between the sample solution and ChatGPT responses}
\label{sec:gap}
To answer the second research question, we first calculated the difference between the judgments of the sample solution (R1.4, R2.4, and R3.6) and the arithmetic mean of ratings of the three bot responses for each question. We did this separately for the two scales (scientific accuracy and linguistic quality). We then subjected this gap metric to an ANOVA with performance assessment as the between-subjects factor (the students only gave ratings from 1 to 6 points for their self-assessment, therefore this factor is modeled as 6-level). 

\begin{figure*}[t!]
    \centering
    \includegraphics[width=\textwidth]{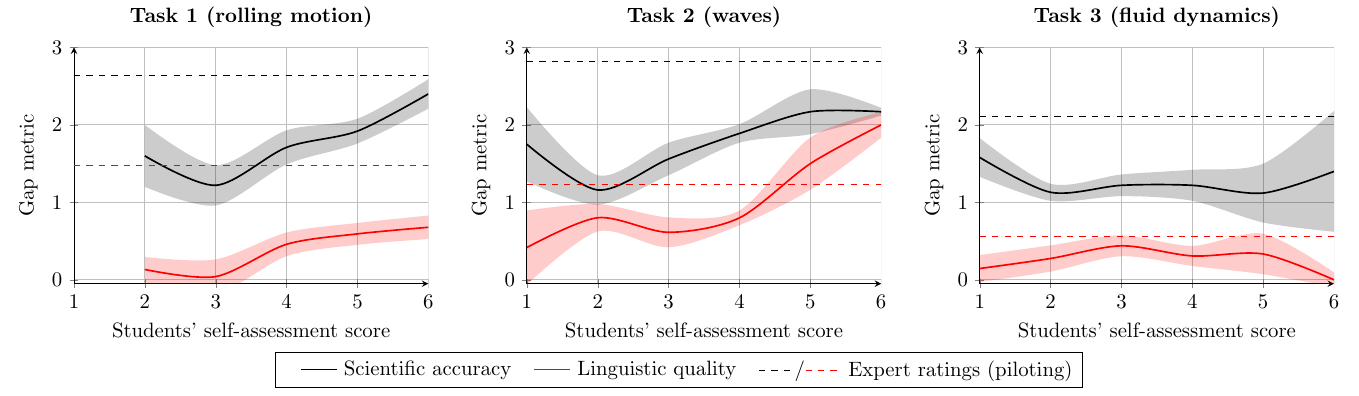}
    \caption{Scientific accuracy- and linguistic quality-related gap metric (distance between the sample solution and the arithmetic mean of the three bot responses) for each of the three questions displayed with the mean (curve) and standard error (shaded area) over the students' self-assessment score (the expected number of points from 0 to 6 in a written exam). The dashed line represents the gap metric of the experts in the pilots, for which there is no self-assessment score, so this is the average of all experts who rated the responses to that question.}
    \label{fig:Distanz}
\end{figure*}

This gap metric can now be used to discuss the effect of the students' self-assessment scores. For each of the three questions, in FIG. \ref{fig:Distanz}, the gap metric is presented over the students' self-assessment scores. For comparison, also the experts' gap metric from the pilots is presented there.

In the first question, we observed a statistically significant main effect of the self-assessment score on the performance gap metric ($F(4,91)=3.42, p=0.007, \eta^2=0.14$), i.e. a higher self-assessment score is correlated with a higher discrepancy in perceived scientific accuracy between the sample solution and ChatGPT responses.  However, we did not find any significant impact on the gap metric related to linguistic quality. In the second question, both gap metrics varied based on the students' self-assessed performance score (performance: $F(5, 88)=2.20, p=0.05, \eta^2=0.11$; linguistic quality: $F(5, 88)=2.52, p=0.04, \eta^2=0.13$). However, in the third question, no such relationship was present.
Moreover, the descriptive data in Fig. \ref{fig:Distanz} show that the experts assess the scientific accuracy of the chatbot responses more differently from the sample solution than the students. The data also show that for questions 1 and 2 (low and medium level of difficulty), students with a higher self-assessment score rate the responses more expert-like than students with a lower self-assessment score (black line in the diagram has a positive slop) while for question 3 (high level of difficulty), the students' rating of the responses is almost independent of their self-assessment score (the black line has almost zero slope). As the students' self-assessment score can carefully be treated as an indicator of the students’ prior knowledge, this demonstrates that with increasing, expert-like prior knowledge, students (as experts) rate the sample solution much better than the faulty ChatGPT responses while students with lower, novice-like prior knowledge rate the faulty ChatGPT responses not that distinctly different than the sample solution.

\subsection{Impact of perceived linguistic quality on scientific accuracy ratings}
\label{sec:impact}
To address the third research question, we restructured the dataset to enable an analysis of covariance (ANCOVA). Specifically, we treated the presented responses as an independent variable with four values (e.g. R1.1, R1.2, R1.4, and R1.6), the rating of scientific accuracy as the dependent variable, and the rating of linguistic quality as a covariate. Here, we present a comparison of the main effect (i.e., the impact of the presented responses on the perceived scientific accuracy) with and without the covariates. The correlations between both scales are $r(384)=.45, p<.01, r(377)=.60, p<.01$, and $r(374)=.47, p<.01$ for questions 1, 2, and 3, respectively.

For the first question, the analysis of covariance shows a significant effect of the response on the assessment of scientific accuracy [$F(3,383)=130, p<.001, \eta^2=0.46$] when corrected for the different perceptions of linguistic quality. Without correction, the effect also occurs and is marginally larger [$F(3,383)=141, p<.001, \eta^2=0.53$]. In the second question, the effect was more pronounced without correction [$F(3, 376)=121, p<.001, \eta^2=0.49$] than with correction [$F(3,376)=87.4, p<.001, \eta^2=0.26$]. In the third question, there was also a substantial difference concerning the effect sizes between both models [$F(3,373)=152, p<.001, \eta^2=0.43$ and $F(3,373)=165, p<.001, \eta^2=0.57$ with and without correction, respectively]. For a better overview, the values for $\eta^2$ are shown in Table~\ref{tab:etasquared}.

In summary, when accounting for differences in the linguistic quality of the responses, the effect of the presented responses on the perceived scientific accuracy still exists, but the effect is less pronounced, as indicated by smaller effect sizes $\eta$. A lower effect size in ANCOVA compared to ANOVA  suggests that controlling for the covariate has reduced the influence of the independent group variable (response type) on the dependent variable (perceived scientific accuracy). This indicates that the covariate plays an important role in explaining the dependent variable, especially in question 2, which is also supported by a high correlation between both scales. This implies that the perceived linguistic quality has a high impact on the perceived scientific accuracy. 

\begin{table}
\caption{Comparison of effect sizes ($\eta^2$) that indicate the impact of the different responses on the perceived scientific accuracy. In model 1, the accuracy ratings are corrected by the perceived linguistic quality (i.e., the linguistic quality was treated as a covariate in the model) whereas in model 2 no such correction was performed. }
\begin{ruledtabular}
\begin{tabular}{p{0.5\columnwidth}p{0.14\columnwidth}p{0.18\columnwidth}}
&\multicolumn{2}{c}{$\eta^2$}\\
&model 1 (ANCOVA) & model 2 (ANOVA)\\\hline
Question 1 (rolling motion)     &0.46  &0.53\\
Question 2 (waves)     &0.26 &0.49\\
Question 3 (fluid dynamics)&0.43&0.57
\end{tabular}
\end{ruledtabular}
\label{tab:etasquared}
\end{table}

\section{Discussion}

In this section, we answer the three research questions from section~\ref{sec:rq} and point out the limitations of the current study.

Regarding the first research question, the results of our study suggest that students assess the linguistic quality of ChatGPT responses on the same level as the sample solutions. The scientific accuracy, however, is generally rated on a lower level than that of the sample solutions. The latter effect is more pronounced for questions of a lower than for a higher level of difficulty. 

One response to the question of the highest level of difficulty was rated on the same level of scientific accuracy as that of the sample solution. It is likely that the good rating of this one response was mostly due to the fact that there weren't any wrong statements in the response, and it gave the impression to answer the question. This is in line with a known limitation of ChatGPT, that it gives "plausible-sounding but incorrect or nonsensical answers" \cite{openai_chatgpt_2022}. 

Regarding the second research question, we found that the gap between students' assessment of the scientific accuracy of ChatGPT responses and sample solutions is significantly lower if the (self-assessed) level of expertise is lower ( Fig.~\ref{fig:Distanz}). This might be explained by the illusion of understanding: If the answer to a question is unknown (which is the case if the self-assessed score is low), plausible-sounding responses sound correct, no matter the actual level of correctness. Therefore, the results of this study add to research about the illusion of understanding (cf. section \ref{sec:illusion}) by giving further evidence that students with lower prior knowledge are more susceptible to inadequate physics conceptions.

Regarding the third research question, the results of our study suggest that the perception of the linguistic quality does impact the perception of scientific accuracy, i.e. the perception that a response is well written might overlay the judgment of scientific accuracy, even though the actual scientific accuracy and linguistic quality are likely not related with each other. This result is particularly interesting regarding the fact that the linguistic quality of ChatGPT responses is already of unmatched quality, whereas the scientific accuracy is still partly questionable (cf. the responses in this study as well as \cite{kortemeyer_could_2023,west_ai_2023,gregorcic_chatgpt_2023}).

The study indicates the risks of using ChatGPT as a student when it is used to find answers to unknown questions. Since ChatGPT does not show any signs of metacognition or a confidence scale for its response, it is up to the user to decide how much the particular response can be trusted. As research about the illusion of understanding shows, students tend to trust plausible sounding responses especially when they foster common false conceptions they hold themselves. The findings from the first part of the survey about students' performance expectations and attitudes towards AI suggest that while they recognize the potential limitations of AI-generated answers, they still consider chatbots to be useful tools for learning with the expectation of receiving reliable information.

At the same time, the study also hints at the potential of ChatGPT as an educational tool, since false ChatGPT responses can be used to educate students, e.g., in an activity like "spot the bot" as suggested by \cite{west_ai_2023}. Research about learning with errors shows that such activities are most beneficial for advanced students and only with an explicit intervention \cite{grose_finding_2007,mason_advanced_2010}. Upon asking at the end of the survey, some students stated that they found this small exercise to be very instructive; some even wished for more exercises of this kind. This demonstrates the potential of using chatbot-generated responses to help students reflect on their understanding and evaluate the scientific accuracy of explanations, providing a unique opportunity to promote metacognition and stimulate students to reflect on their understanding and reasoning. It was pointed out by \cite{gregorcic_chatgpt_2023} that using responses from ChatGPT might be useful in pre-service teacher education to learn and recognize problematic argumentation without being distracted by grammatical or stylistic issues.

The current study had some limitations. 
The most important factor which influences the outcome of this study is the creation of the responses provided by ChatGPT as well as the sample solution. ChatGPT generally rarely gives the same response to the same question, hence in principle, there would be an infinite amount of possible ChatGPT responses. We tried to account for this fact by giving the students three different responses per question which we already chose out of a collection of five different generated responses. In these five different responses we already saw some repetition in explanation patterns (cf. Table~\ref{tab:qandr}), but this does not mean that in more iterations we would not find new patterns. However, this randomness factor of ChatGPT cannot be influenced by the researchers, as the responses given by the chatbot are not deterministic.

Besides this limitation, students had to rate the responses based on five criteria which were not further specified, therefore no gauging of the rating took place. We did not use detailed descriptions of the criteria to stay within the time frame and to avoid unnecessary cognitive load on the students. This unguided rating of the answers might have caused judgment errors, such as a bias toward extreme values or the center. Also, we have not specified in detail what we mean by the term "sample solution", so we can only assume that students judged the \textit{suitability as sample solution} as sample solution for themselves. In future work, these judgment errors could be avoided by the implementation of more detailed scales or rating manuals for the cost of a longer survey duration and a higher cognitive load for the students.

We presented the responses in a randomized order to avoid a rating bias for the first presented response. This, however, lead to the fact that students saw the sample solution at different positions. If students did not know the correct response, seeing the correct response might have influenced the rating of responses presented afterward. 

As already pointed out in section~\ref{sec:survey}, the prior knowledge of the students was assessed by a self-assessment score. This is only a very distant proxy of assessing the expertise on a question. We chose this method to keep the survey to be feasible within a limited time frame and to not bias the students' ratings too much. Some of the students might have started to craft a solution in their minds which they compared the presented responses to, especially in the case of the first question, where the students took on average 60\,s (SD = 33\,s) to do the self-assessment. For the questions 2 and 3 this time was lower (43\,s, (SD = 28\,s) and 46\,s (SD = 35\,2), respectively) even though the questions were more difficult, so they didn't spend much time crafting their own sample solution. 
Regarding the validity of the self-assessment as a proxy for expertise, as shown in section~\ref{sec:self}, students rated themselves highest for the question with the lowest level of difficulty, a bit lower for the question with the medium level of difficulty, and lowest for the question with the highest level of difficulty. This is a hint that the self-assessment score is not completely uncorrelated with the actual level of expertise. Since the means of the self-assessment score did not differ significantly between first- and second-year students, the intended level of difficulty can be assumed to hold for the two groups which is why we treated both samples as one group in the further analysis.

Finally, this study was conducted using responses created based on the model GPT-3.5. By the time of writing, an enhanced model, GPT-4, was already published. It is possible that the responses this model gives would be more accurate \cite{west_advances_2023}. While it is true that new advancements in the field may lead to more accurate results in the future, the main findings and conclusions of our study should still hold value for researchers and practitioners alike,  providing a valuable baseline for comparison and further research in this area.

\section{Conclusion and outlook}

Our study shows that students can adequately evaluate the scientific quality of language-wise comparable ChatGPT responses and sample solutions for qualitative physics questions as far as their prior knowledge is sufficient. However, if the students' prior knowledge is limited, it becomes much more challenging for the students to distinguish between correct and complete sample-solution-like responses and ChatGPT responses that are not incorrect but do not address the key aspect of the problem.

Thus, instructors also in physics studies will need to educate their students in adequately evaluating and responsibly using ChatGPT responses for their studies. For this, ChatGPT can serve as an educational tool in analogy to \textit{worked examples} that provides students with unlimited responses to all physics questions that can and should be analyzed regarding the factual correctness and degree of completeness. Instructors need to guide their students and teach them how to conduct this evaluation process based on the knowledge taught in the study course program and by further research (e.g., on the Internet or standard textbooks).

In perspective, research on ChatGPT in physics education needs to be continued. Based on our findings in this survey, we would particularly propose two research questions: First, one needs to investigate to which extent students trust ChatGPT responses in comparison to traditional textbooks or other Internet resources including forums, and accordingly if ChatGPT could substitute traditional teaching and learning resources from the student’s point of view. This could be tested in a similar survey as described here by additionally just masking the pretended origin (e.g., ChatGPT, textbook, website, chat forum, ...) of the responses randomly. Second, in preparation for our study, we experienced that ChatGPT comes up with very different (wrong or misleading) explanations for identical physics questions, revealing different misconceptions. Thus, a systematic analysis of ChatGPT responses could be conducted to identify the (probably language-dependent) "misconceptions" of ChatGPT and to check whether they are in accordance with known students' preconceptions already described in the literature. This comparison would allow an appraisal of whether ChatGPT is connectable to the students' preconceptions or whether it produces new preconceptions which are rare among students.

ChatGPT itself is not the first and not the last generative AI tool that exists. Already at the time of writing the paper, there is the direct successor of the version the authors used in the current study available with further developments in sight. An interesting development would be a tool which can calculate confidence levels of responses a certain AI gives based on its training. This tool could be used to address concerns about a too authoritative response style. 

The current study is  still very early in its field and hence of an exploratory character. Further research in this field is highly encouraged and will continue to give valuable insights into how learners use and interact with AI tools such as ChatGPT and which skill sets will be needed to do so in a responsible and meaningful way.

\section*{Author Contributions}
All authors contributed to all stages of this research, including data collection, analysis, and the writing of the present paper. All authors have read and approved the manuscript.

\section*{Ethics statement}
Participation in the survey was not mandatory and their participation did not influence the examination. Students gave their consent to the use of their survey responses in our research.
\begin{acknowledgements}
The authors would like to thank the students and physics faculty who took part in validating and piloting the survey instruments.
MD received funding from the Wolfgang Gentner Programme of the German Federal Ministry of Education and Research (grant no. 13E18CHA).
\end{acknowledgements}

\clearpage
\appendix

\section{Survey instrument}
\begin{longtable*}{p{0.5\textwidth}|p{.5\textwidth}}
\caption{Survey questions and translations of the survey instrument described in section \ref{sec:survey}}\\
\hline\hline
\textbf{German original} & \textbf{English translation}\\\hline\hline
In welchem Studiengang sind Sie eingeschrieben?	&What is your field of study?\\
\textit{Physik Mono-Bachelor, Physik 2FB Lehramt – Mein Zweifach ist:, Physik 2FB (nicht Lehramt) – Mein Zweitfach ist:, Sonstiges – Mein Studiengang ist:}	&\textit{Physics major, Physics education – My second subject is:, Physics double major – My second subject is:, Other – My field of study:}\\\hline
In welchem Fachsemester in Physik befinden Sie sich aktuell?&	Please select your current semester:\\
\textit{1. Fachsemester, 2. Fachsemester, 3. Fachsemester, \textgreater{}3. Fachsemester	}&\textit{1. Semester, 2. Semester, 3. Semester, \textgreater{}3. Semester}\\\hline
Welchem Geschlecht fühlen Sie sich zugehörig?	&What gender do you identify as?\\
\textit{Weiblich, männlich, divers, keine Angabe}	&\textit{female, male, diverse, prefer not to say}\\\hline
Wie sehr stimmen Sie folgenden Aussagen zu?	&To what extent do you agree with the following statements?\\
\textit{Stimme gar nicht zu, stimme nicht zu, teils/teils, stimme zu, stimme voll zu, weiß nicht/keine Angabe}&	\textit{Strongly disagree, Disagree, Neither agree nor disagree, Agree, Strongly agree, Don’t know/prefer not to say}\\
-- Durch einen Chatbot kann ich präzise Antworten erhalten.	& -- A chatbot allows me to get accurate answers.\\
-- Antworten eines Chatbots sind nicht immer korrekt.&	-- Answers from a chatbot are not always correct.\\
-- Künstliche Intelligenz (KI) ist für mein Studium nützlich.&	-- Artificial intelligence (AI) is useful for my studies.\\\hline
Haben Sie schon einmal von ChatGPT gehört?	&Have you heard of ChatGPT before?\\
\textit{Ja/Nein}&	\textit{Yes/No}\\\hline
Wie oft haben Sie schon ChatGPT zur Bearbeitung von Übungszetteln in Ihrem Physikstudium genutzt?	&How many times have you used ChatGPT to work on exercises in your physics studies so far?\\
\textit{Noch nie, einmal, zwei- bis fünfmal, öfter als fünfmal}	&\textit{Never, once, 2-5 times, more than 5 times}\\\hline
Wie oft haben Sie schon ChatGPT für andere Dinge als Übungszettel (z. B. privat) genutzt?	&How many times have you used ChatGPT for things other than exercises (e.g., privately)?\\
\textit{Noch nie, einmal, zwei- bis fünfmal, öfter als fünfmal}&	\textit{Never, once, 2-5 times, more than 5 times}\\\hline
Informationen zur Umfrage\newline
ChatGPT ist ein Bot, der derzeit in vielen Bereichen der Gesellschaft diskutiert wird und auch im Bereich der Physik-Lehre genutzt wird. Der Bot produziert unter Umständen sehr unterschiedliche Antworten auf dieselbe Frage.\newline
Nachfolgend werden Ihnen drei Physik-Erkläraufgaben angezeigt. Lesen Sie sich diese in Ruhe durch. Sie sollen die Aufgaben nicht selbst bearbeiten, sondern nur eine schnelle Selbsteinschätzung geben, wie Sie bei der jeweiligen Aufgabe in einer Klausur abschneiden würden.\newline
Anschließend werden je Aufgabe nacheinander vier Lösungen angezeigt, die von ChatGPT erstellt wurden. Sie können sich die Aufgabe dabei immer noch einmal ansehen.\newline
Dann sollen Sie die Lösungen anhand folgender fünf Kriterien einschätzen: fachliche Korrektheit, Grad der Vollständigkeit, Verständlichkeit, sprachliche Qualität und Eignung als Musterlösung.\newline
Es geht dabei nicht darum, die Lösungen im Vergleich zueinander zu bewerten, sondern jede Lösung für sich anhand des ersten Eindrucks einzuschätzen.&
Survey information\newline
ChatGPT is a bot that is currently being discussed in many areas of society and is also being used in the field of physics education. The bot may produce very different answers to the same question.\newline
Below, you will see three physics explanation questions. Read through them at your leisure. You are not supposed to work on the tasks yourself, but only give a quick self-assessment of how you would perform on each task in an exam.\newline
Afterwards, four solutions per task are displayed one after the other, which were created by ChatGPT. Meanwhile, you can always have another look at the task.\newline
Then, you have to evaluate the solutions based on the following five criteria: technical correctness, degree of completeness, comprehensibility, linguistic quality and suitability as a sample solution.\newline
It is not a matter of evaluating the solutions in comparison to each other, but rather of evaluating each solution individually based on the first impression.\\\hline
Ohne die Aufgabe selbst zu lösen: Schätzen Sie ein, wie viele Punkte Sie in einer Klausur bei dieser Aufgabe erhalten hätten (0-6 Punkte)?&	Without solving the task yourself: Estimate how many points you would have received on this task in a written exam (0-6 points)\\\hline
Bewerten Sie die Antwort bitte anhand folgender Kriterien:	&Please rate the answer based on the following criteria:\\
\textit{Sehr niedrig, niedrig, mittel, hoch, sehr hoch}&	\textit{Very low, low, medium, high, very high}\\
-- Fachliche Korrektheit	& -- Factual correctness\\
-- Grad der Vollständigkeit	& -- Degree of completeness\\
-- Verständlichkeit	& -- Comprehensibility\\
-- Sprachliche Qualität	& -- Quality of the language\\
-- Eignung als Musterlösung&	-- Suitability as a sample solution\\\hline
Abschlussfrage\newline
Möchten Sie uns abschließend noch etwas zur Umfrage mitteilen? Dann können Sie das hier tun...	&
Final question\newline 
Would you like to tell us something concerning the survey in conclusion? Then you can do so here...\\\hline\hline
\label{tab:instrument-translation}
\end{longtable*}

\clearpage
\section{Questions and responses}

\begin{longtable*}{p{.05\textwidth}|p{.36\textwidth}|p{.36\textwidth}|p{.20\textwidth}}
\caption{Questions used in the survey instrument (denoted by Q1, Q2, and Q3) and responses generated by ChatGPT (R1.1, R1.2, R1.3, R1.5, R1.6, R2.1, R2.2, R2.3, R2.5, R2.6, R3.1, R3.2, R3.3, R3.4, R3.5) and the sample solution written by the authors (R1.4, R2.4, R3.6). For a detailed description on the creation of the questions and responses see section \ref{sec:survey}. In the case of the second question, ChatGPT gave rather long responses. To stay concise, it was asked to shorten these to one paragraph. The table gives the long responses, indicated as "init" in the first column, as well as the final responses.}\\
\hline\hline
\textbf{Q/R}            & \textbf{German Original}& \textbf{English translation}& \textbf{Comment}      \\\hline\hline

Q1       & Beim Vorlesungsexperiment ”Wettrennen der Zylinder am Hang“ rollten   ein Vollzylinder (Radius R, Länge L, Masse M) und ein Hohlzylinder   (Außenradius R, Länge L, Masse M) eine schiefe Ebene hinab. Zu Beginn waren   beide Zylinder in Ruhe und befanden sich auf gleicher Ausgangshöhe. Welcher   Zylinder erreichte als erstes das untere Ende der Ebene? Begründen Sie Ihre   Antwort!                                                                                                                                                                                                                                                                                                                   & In the lecture experiment   "Race of cylinders on a slope", a solid cylinder (radius R, length   L, mass M) and a hollow cylinder (outer radius R, length L, mass M) rolled   down an inclined plane. At the beginning, both cylinders were at rest and   were at the same initial height. Which cylinder reached the bottom end of the   plane first? Give reasons for your answer!                                                                                                                                                                                                                                                                                              & Low level of difficulty (repetition of   an already known question).   \\\hline
R1.1 & Der Hohlzylinder wird als erstes das untere Ende der Ebene erreichen,   da er ein niedrigeres Trägheitsmoment hat, was bedeutet, dass er weniger   Widerstand gegen die Rotationsbewegung hat. Dies macht es dem Hohlzylinder   einfacher, die Schräge hinabzurollen und erreicht daher schneller als der   Vollzylinder das untere Ende.                                                                                                                                                                                                                                                                                                                                                                       & The hollow cylinder will reach the   bottom end of the incline first because it has a lower moment of inertia, meaning it will have less resistance to rotational motion. This makes it easier for   the hollow cylinder to roll down the incline, allowing it to reach the   bottom  faster than the solid cylinder.                                                                                                                                                                                                                                                                                                                                                        & Only ChatGPT response with obviously   wrong result stating that the hollow cylinder reaches the end of the inclined   plane quicker than the solid cylinder. The quality of the answer is judged   similarly to responses 1.3 and 1.5 in the pilots. In this case, the response given by ChatGPT was initially given in English and the German answer was generated by asking to translate the response to German. \\*
R1.2 & Der Vollzylinder erreicht als erstes das untere Ende der Ebene. Dies   ist aufgrund des höheren Trägheitsmoments des Vollzylinders im Vergleich zum   Hohlzylinder der Fall. Da das Trägheitsmoment proportional zur Masse und dem   Radius des Zylinders ist, ist das Trägheitsmoment des Vollzylinders größer   als das des Hohlzylinders. Dies führt dazu, dass der Vollzylinder eine   geringere Neigung bei gleicher Anstoßkraft aufweist, wodurch er schneller die   schiefe Ebene hinabrollt.                                                                                                                                                                                                            & The solid cylinder is the first to   reach the bottom end of the plane. This is the case due to the higher moment   of inertia of the solid cylinder compared to the hollow cylinder. Since the   moment of inertia is proportional to the mass and radius of the cylinder, the   moment of inertia of the solid cylinder is greater than that of the hollow   cylinder. This results in the solid cylinder having a lower inclination for   the same impact force, causing it to roll down the inclined plane faster.                                                                                                                                                            & Best rated ChatGPT response in the   pilots and therefore chosen for the main survey.   \\
R1.3 & Der Vollzylinder wird als erstes das untere Ende der Ebene erreichen.   Dies liegt daran, dass bei gleicher Masse und gleicher Länge die Trägheit des   Vollzylinders größer ist als die des Hohlzylinders, da dessen Material in   einer größeren Entfernung zur Rotationsachse angeordnet ist. Da Trägheit und   Zeit in einem solchen Experiment proportional sind, wird der Vollzylinder   länger brauchen, um das Ende der Ebene zu erreichen.                                                                                                                                                                                                                                                             & The solid cylinder will reach the   lower end of the plane first. This is because, for the same mass and length,   the inertia of the solid cylinder is greater than that of the hollow cylinder   because its material is located at a greater distance from the axis of   rotation. Since inertia and time are proportional in such an experiment, the   solid cylinder will take longer to reach the end of the plane.                                                                                                                                                                                                                                                         & Excluded based on the pilot   results because similarly rated as responses 1.1 and 1.5 but with a similar   result as the other chosen responses.   \\
R1.4 & Der Vollzylinder erreicht als erstes das untere Ende der Ebene, da er   ein kleineres Trägheitsmoment hat. Die gleiche anfängliche potenzielle   Energie wird beim Herabrollen der schiefen Ebene nach der Energieerhaltung in   kinetische Translationsenergie und Rotationsenergie umgewandelt. Da das   Trägheitsmoment für den Vollzylinder kleiner als für den Hohlzylinder ist (da   seine Masse weiter nach Innen verteilt ist), muss folglich der   Rotationsanteil an der Energie für den Vollzylinder kleiner sein und der   Translationsanteil höher sein. Somit ist die Geschwindigkeit des   Vollzylinders zu jedem Zeitpunkt höher.                                                               & The solid cylinder is the first to   reach the bottom of the plane because it has a smaller moment of inertia. The   same initial potential energy is converted into kinetic translational energy   and rotational energy when rolling down the inclined plane according to the   conservation of energy. Since the moment of inertia for the solid cylinder is   smaller than for the hollow cylinder (since its mass is distributed further   inward), consequently the rotational part of the energy for the solid   cylinder must be smaller and the translational portion must be higher. Thus,   the velocity of the solid cylinder is higher at all times.                 & Sample solution; "zu jedem   Zeitpunkt“ (translation: “at all times”) was added to the last sentence based   on the pilots.     \\
R1.5 & Es ist wahrscheinlich, dass der Vollzylinder zuerst das untere Ende   erreichen wird. Dies ist aufgrund seiner höheren Trägheit zu erklären. Da ein   Hohlzylinder einen höheren Schwerpunkt hat als ein Vollzylinder, wird er eher   den Hang hinuntergleiten und aufgrund seiner geringeren Trägheit weniger   gegenüber Störungen wie etwa Unebenheiten anfällig sein. Der Vollzylinder hingegen   hat aufgrund seiner größeren Masse und niedrigeren Schwerpunktlage eine   höhere Trägheit und wird daher weniger anfällig für Störungen sein, wodurch   er wahrscheinlich als erstes das untere Ende erreicht.                                                                                            & It is likely that the solid   cylinder will reach the lower end first. This can be explained due to its   higher inertia. Since a hollow cylinder has a higher center of gravity than a   solid cylinder, it will be more likely to slide down the slope and less   susceptible to disturbances such as bumps due to its lower inertia. The solid   cylinder, on the other hand, has a higher inertia due to its greater mass and   lower center of gravity, and will therefore be less susceptible to   disturbances, due to which it is likely to reach the bottom end first.                                                                                                   & Excluded based on the pilot   results because similarly rated as responses 1.1 and 1.3 but with a similar   result as the other chosen responses. \\
R1.6 & Es ist nicht möglich zu sagen, welcher Zylinder als erstes das untere   Ende der Ebene erreicht, ohne weitere Informationen über die schiefe Ebene,   die Reibung und andere Faktoren. Beide Zylinder haben die gleiche Masse, aber   unterschiedliche Formen, die ihre Beschleunigung beeinflussen können. Eine   genaue Prognose erfordert eine Analyse der kinematischen und dynamischen   Prinzipien, die auf die Zylinder wirken.                                                                                                                                                                                                                                                                          & It is not possible to say which   cylinder will reach the bottom of the plane first without more information   about the inclined plane, friction, and other factors. Both cylinders have   the same mass, but different shapes that can affect their acceleration. An   accurate prediction requires an analysis of the kinematic and dynamic   principles acting on the cylinders.                                                                                                                                                                                                                                                                                              & Worst rated ChatGPT response in the   pilots and therefore chosen for the main survey.  \\\hline\hline
Q2       & Bei einem Experiment werden ein Lautsprecher und davor zwei brennende   Kerzen hintereinander positioniert. Die beiden Kerzen haben einen Abstand von   einem Meter zueinander. Der Lautsprecher erzeugt einen kontinuierlichen Ton   der Frequenz 170 Hz. Die Schallgeschwindigkeit beträgt in diesem Fall 340   m/s. Beschreiben Sie das Verhalten der beiden Kerzenflammen zunächst einzeln   und dann im Vergleich zueinander. Begründen Sie Ihre Antwort!                                                                                                                                                                                                                                                  & In an experiment, a loudspeaker   and, in front of it, two burning candles are positioned behind each other. The   two candles have a distance of one meter from each other. The loudspeaker   produces a continuous sound with a frequency of 170 Hz. The speed of sound in   this case is 340 m/s. Describe the behavior of the two candle flames, first   individually and then in comparison with each other. Give reasons for your   answer!                                                                                                                                                                                                                                 & Medium level of difficulty   (application of knowledge learned just a few weeks ago).            \\\hline
R2.1 (init) & 
Die beiden   Kerzenflammen werden durch den Schall beeinflusst, da sich durch den Schall   die Luftbewegungen ändern. Da der Ton eine kontinuierliche Frequenz hat, wird   die Luftbewegung regelmäßig und die Flammen werden daher ebenfalls regelmäßig   hin und her bewegt.\newline Einzeln   betrachtet wird jede Flamme unabhängig von der anderen beeinflusst. Die   Intensität der Flammenbewegung ist abhängig von der Intensität des Schalls   und dem Abstand zum Lautsprecher.\newline Im Vergleich   zueinander werden die Flammen unterschiedlich beeinflusst, da sie   unterschiedlich weit vom Lautsprecher entfernt sind. Da sich die   Schallgeschwindigkeit konstant und gleichmäßig ausbreitet, wird die Flamme,   die näher am Lautsprecher ist, stärker beeinflusst als die, die weiter weg   ist. Daher werden die Flammen unterschiedlich große Bewegungen ausführen.&
The two   candle flames are affected by the sound because the sound changes the air   movements. Since the sound has a continuous frequency, the air movement   becomes regular and the flames are therefore also regularly moved back and   forth.\newline Individually,   each flame is affected independently of the other. The intensity of the flame   movement depends on the intensity of the sound and the distance to the   speaker.\newline Compared to   each other, the flames are affected differently because they are at different   distances from the loudspeaker. Since the speed of sound propagates   constantly and uniformly, the flame that is closer to the speaker will be   affected more than the one that is farther away. Therefore, the flames will   move at different rates. &
Initial response from ChatGPT. To get a more concise answer, ChatGPT was asked to summarize the answer in one paragraph which lead to response R2.1\\

R2.1 & Bei einem Experiment mit einem Lautsprecher, der einen   kontinuierlichen Ton der Frequenz 170 Hz erzeugt, und zwei brennenden Kerzen,   die einen Abstand von einem Meter zueinander haben, werden die Kerzenflammen   durch den Schall beeinflusst. Jede Flamme wird unabhängig von der anderen beeinflusst,   aber im Vergleich zueinander werden sie unterschiedlich beeinflusst, da die   nähere Flamme stärker beeinflusst wird als die entferntere.                                                                                                                                                                                                                                                      & In an experiment with a   loudspeaker producing a continuous sound of frequency 170 Hz and two burning   candles, which have a distance of one meter from each other, the candle   flames are affected by the sound. Each flame is affected independently of the   other, but compared to each other they are affected differently, as the   closer flame is affected more than the more distant one.                                                                                                                                                                                                                                                                             & Best rated ChatGPT response regarding   the scientific accuracy in the pilots and therefore chosen for the main   survey.  \\

R2.2 (init) &
Wenn ein   kontinuierlicher Ton der Frequenz 170 Hz vom Lautsprecher erzeugt wird, kann   man bei den beiden brennenden Kerzen eine Interferenz des Schalls beobachten.   Die beiden Kerzen stehen einen Meter voneinander entfernt, was einem halben   Wellenlängenabstand entspricht. Daher ist das Verhalten der Flammen   periodisch, sie werden größer und kleiner, je nachdem ob die Schallwellen in   Phasenaddition oder -subtraktion stehen.\newline Im Vergleich   zueinander kann man beobachten, dass die Flammen bei einem Zeitpunkt   gleichzeitig größer werden und bei einem anderen Zeitpunkt gleichzeitig   kleiner werden. Dies ist ein Indikator für eine Interferenz im Schallfeld,   bei der Phasenaddition und -subtraktion stattfindet.\newline Dieses   Phänomen wird durch den Dopplereffekt verursacht, bei dem sich die Frequenz   des Tons ändert, je nachdem ob sich die Schallquelle auf den Beobachter zu   oder von ihm weg bewegt. Da sich die Schallquelle (der Lautsprecher) nicht   bewegt, sind beide Flammen gleichzeitig Schallquellen und Schallempfänger,   wodurch eine Interferenz im Schallfeld entsteht. & 
When a   continuous sound of frequency 170 Hz is produced by the loudspeaker, an   interference of the sound can be observed at the two burning candles. The two   candles are one meter apart, which corresponds to half a wavelength distance.   Therefore, the behavior of the flames is periodic, they become larger and   smaller depending on whether the sound waves are in phase addition or   subtraction.\newline In comparison   with each other, it can be observed that the flames become larger at one time   and smaller at another time simultaneously. This is an indicator of   interference in the sound field, where phase addition and subtraction occurs.\newline This   phenomenon is caused by the Doppler effect, in which the frequency of the   sound changes depending on whether the sound source is moving toward or away   from the observer. Since the sound source (the loudspeaker) does not move,   both flames are sound sources and sound receivers at the same time, causing   interference in the sound field. &
Initial response from ChatGPT. To get a more concise answer, ChatGPT was asked to summarize the answer in one paragraph which lead to response R2.2\\

R2.2 & Bei einem Experiment mit einem kontinuierlichen Ton der Frequenz 170   Hz und einem Abstand von einem Meter zwischen zwei brennenden Kerzen vor   einem Lautsprecher, kann man eine Interferenz im Schallfeld beobachten. Die   Flammen werden größer und kleiner in einem periodischen Muster, da sie   gleichzeitig Schallquellen und Empfänger sind. Dies ist durch den   Dopplereffekt verursacht, bei dem sich die Frequenz des Tons ändert, je nach   Bewegung der Schallquelle.                                                                                                                                                                                                                          & In an experiment with a continuous   sound of frequency 170 Hz and a distance of one meter between two burning   candles in front of a loudspeaker, one can observe interference in the sound   field. The flames become larger and smaller in a periodic pattern as they are   both sound sources and receivers. This is caused by the Doppler effect, where   the frequency of the sound changes depending on the movement of the sound   source.                                                                                                                                                                                                                               & The only ChatGPT response which   refers to the Doppler effect and does not refer to phases. Besides that,   similar quality rated in the pilots as for responses 2.3 and 2.5, thus chosen   for the main survey.\\

R2.3 (init)&
Wenn der   Lautsprecher einen kontinuierlichen Ton der Frequenz 170 Hz ausstrahlt, kann   es zu einer Interferenz der Schallwellen vor den beiden Kerzen kommen. Eine   Schallwelle wird von dem Lautsprecher ausgestrahlt und trifft auf die beiden   Kerzen. Die Schallwelle, die auf die erste Kerze trifft, wird gestreut und   wird als neue Schallwelle zurück an die zweite Kerze gesendet. Wenn die Phase   der Schallwelle, die von der ersten Kerze zurückgesendet wird, genau der   Phase der Schallwelle entspricht, die auf die zweite Kerze trifft, dann   werden die beiden Wellen überlagert und es kann zu einer Verstärkung oder   einer Dämpfung des Tones kommen, was sich auf die Größe der Kerzenflammen   auswirken kann.\newline Da der   Abstand zwischen den beiden Kerzen einen Meter beträgt, kann es vorkommen,   dass sich die Phasen der beiden Schallwellen ändern, was zu einer Veränderung   des Tones und somit auch zu einer Veränderung der Größe der beiden   Kerzenflammen führen kann.\newline Im Vergleich   zueinander kann es vorkommen, dass die Größe der einen Kerzenflamme größer   ist als die der anderen Kerzenflamme, je nachdem, wie sich die Phasen der   beiden Schallwellen ändern. Es kann auch vorkommen, dass die Größe beider   Kerzenflammen im gleichen Maße verändert wird.\newline Zusammenfassend   kann man sagen, dass das Verhalten der beiden Kerzenflammen durch die   Interferenz der Schallwellen, die von dem Lautsprecher ausgestrahlt werden,   bestimmt wird. Je nach Phasenlage der beiden Schallwellen kann es zu einer   Verstärkung oder Dämpfung des Tones kommen, was sich auf die Größe der   Kerzenflammen auswirkt. & 
If the   loudspeaker emits a continuous sound of the frequency 170 Hz, an interference   of the sound waves in front of the two candles can occur. A sound wave is   emitted from the loudspeaker and hits the two candles. The sound wave that   hits the first candle is scattered and is sent back to the second candle as a   new sound wave. If the phase of the sound wave that is sent back from the   first candle is exactly the same as the phase of the sound wave that hits the   second candle, then the two waves will be superimposed and there may be   amplification or attenuation of the sound, which can affect the size of the   candle flames.\newline Since the   distance between the two candles is one meter, it can happen that the phases   of the two sound waves change, which can lead to a change in the tone and   therefore to a change in the size of the two candle flames.\newline Compared to   each other, it may happen that the size of one candle flame is larger than   that of the other candle flame, depending on how the phases of the two sound   waves change. It can also happen that the size of both candle flames is   changed to the same extent.\newline In summary,   the behavior of the two candle flames is determined by the interference of   the sound waves emitted by the speaker. Depending on the phase position of   the two sound waves, the sound may be amplified or attenuated, which affects   the size of the candle flames.&

Initial response from ChatGPT. To get a more concise answer, ChatGPT was asked to summarize the answer in one paragraph which lead to response R2.3\\
R2.3 & Das Verhalten der beiden Kerzenflammen wird durch die Interferenz der   Schallwellen, die von dem Lautsprecher erzeugt werden, bestimmt. Die Größe   der Kerzenflammen kann sich verändern, je nach Phasenlage der beiden   Schallwellen. Eine Überlagerung der Wellen kann zu einer Verstärkung oder   Dämpfung des Tones führen, was wiederum Auswirkungen auf die Größe der   Kerzenflammen hat.                                                                                                                                                                                                                                                                                                             & The behavior of the two candle   flames is determined by the interference of the sound waves generated by the   loudspeaker. The size of the candle flames can change depending on the phase   position of the two sound waves. Superposition of the waves can lead to   amplification or attenuation of the sound, which in turn affects the size of   the candle flames.                                                                                                                                                                                                                                                                                                        & Excluded based on the pilot   results because similarly rated as responses 2.2 and 2.5 but with a similar   explanation as the other chosen responses. \\
R2.4 & Die Kerzen flackern aufgrund der Schallwelle hin und her, da die   Schallwelle als Longitudinalwelle den Luftdruck an der Kerze periodisch   verändert, wodurch die Flamme periodisch vom Lautsprecher weg und zum   Lautsprecher hinbewegt wird. Die beiden Kerzen befinden sich im Abstand einer   halben Wellenlänge zueinander, dadurch ist ihre Bewegung genau antiparallel   zueinander. Das heißt, wenn eine Flamme vom Lautsprecher weg steht, steht die   andere zum Lautsprecher hin und andersherum.                                                                                                                                                                                                 & The candles flicker back and forth   due to the sound wave, since the sound wave as a longitudinal wave   periodically changes the air pressure at the candle, causing the flame to   periodically move away from the loudspeaker and toward the loudspeaker. The   two candles are at a distance of half a wavelength from each other, thus   their motion is exactly antiparallel to each other. This means that when one   flame is away from the speaker, the other is toward the speaker and vice   versa.                                                                                                                                                                   & Sample solution    \\
R2.5 (init)&
Wenn ein   kontinuierlicher Ton mit einer Frequenz von 170 Hz von einem Lautsprecher   erzeugt wird und vor ihm zwei brennende Kerzen in einem Abstand von einem   Meter zueinander positioniert sind, beobachtet man Interferenzmuster.\newline Einzeln   betrachtet wird jede Kerzenflamme von der Schallwelle beeinflusst, wodurch   sie auf- und abgebogen wird. Dies geschieht aufgrund der Druckunterschiede,   die durch die Schallwelle verursacht werden.\newline Im Vergleich   zueinander werden die Flammen jedoch unterschiedlich beeinflusst. Da sich die   Schallwellen von beiden Kerzen ausbreiten und in der Mitte des Abstands   zwischen ihnen kollidieren, entstehen Interferenzmuster, die auf den   Phasenunterschieden der Wellen beruhen. Wenn sich die Wellen in der Mitte   konstruktiv überlagern, wird die Flamme höher, und wenn sie sich destruktiv   überlagern, wird sie niedriger.\newline Das Verhalten   der Flammen hängt von der Schallfrequenz und dem Abstand zwischen den Kerzen   ab. In diesem speziellen Fall mit einer Frequenz von 170 Hz und einem Abstand   von einem Meter kann man regelmäßige, pulsierende Interferenzmuster   beobachten. & 
When a   continuous sound with a frequency of 170 Hz is produced by a loudspeaker and   two burning candles are positioned in front of it at a distance of one meter   from each other, interference patterns are observed.\newline Individually,   each candle flame is affected by the sound wave, causing it to bend up and   down. This occurs due to the pressure differences caused by the sound wave.\newline However, when   compared to each other, the flames are affected differently. As the sound waves   propagate from both candles and collide in the middle of the distance between   them, interference patterns are created based on the phase differences of the   waves. When the waves constructively overlap in the middle, the flame becomes   higher, and when they destructively overlap, it becomes lower.\newline The behavior   of the flames depends on the frequency of sound and the distance between the   candles. In this particular case with a frequency of 170 Hz and a distance of   one meter, regular pulsating interference patterns can be observed.&

Initial response from ChatGPT. To get a more concise answer, ChatGPT was asked to summarize the answer in one paragraph which lead to response R2.5\\
R2.5 & Wenn ein kontinuierlicher Ton mit einer Frequenz von 170 Hz vor zwei   brennenden Kerzen in einem Abstand von einem Meter erzeugt wird, beobachtet   man Interferenzmuster, die durch die Überlagerung der Schallwellen entstehen.   Jede Kerzenflamme wird von der Schallwelle beeinflusst und auf- und   abgebogen, aber im Vergleich zueinander werden sie unterschiedlich   beeinflusst, da die Schallwellen kollidieren und Interferenzmuster   verursachen, die von den Phasenunterschieden der Wellen abhängen. Dies führt   zu regelmäßigen, pulsierenden Interferenzmustern.                                                                                                                           & When a continuous sound with a   frequency of 170 Hz is generated in front of two burning candles at a   distance of one meter, one observes interference patterns caused by the   superposition of the sound waves. Each candle flame is affected by the sound   wave and bends up and down but compared to each other they are affected   differently because the sound waves collide and cause interference patterns   that depend on the phase differences of the waves. This results in regular,   pulsating interference patterns.                                                                                                                                          & Excluded based on the pilot   results because similarly rated as responses 2.2 and 2.3 but with a similar   explanation as the other chosen responses. \\

R2.6 (init)&
Die beiden   Kerzenflammen reagieren auf den Ton des Lautsprechers durch Interferenz. Da   der Ton eine feste Frequenz hat, bleibt auch die Phasenlage zwischen den   beiden Flammen konstant, was dazu führt, dass sich die Flammen mal verstärken   und mal abschwächen.\newline Die einzelnen   Flammen reagieren auf den Ton durch Veränderungen ihrer Größe, da sich die   Schallwellen auf die Flammen auswirken. Da jedoch jede Flamme eine   unterschiedliche Distanz zum Lautsprecher hat, erreicht die Schallwelle die   eine Flamme früher als die andere. Daher kann es zu Phasendifferenzen   zwischen den beiden Flammen kommen, die zu Interferenz führen.\newline Im Vergleich   zueinander kann es zu Phasenüberlagerungen kommen, wodurch sich die Flammen   mal verstärken und mal abschwächen. Je nach Phasendifferenz kann eine Flamme   größer oder kleiner sein als die andere. Da die Phasendifferenz aufgrund des   festen Tonfrequenz ändert, ändert sich auch das Verhältnis der beiden Flammen   ständig. & 
The two   candle flames react to the sound of the loudspeaker by interference. Since   the sound has a fixed frequency, the phase position between the two flames   also remains constant, which causes the flames to sometimes strengthen and   sometimes weaken.\newline The individual   flames respond to the sound by changing their size as the sound waves affect   the flames. However, since each flame has a different distance from the   speaker, the sound wave reaches one flame earlier than the other. Therefore,   phase differences can occur between the two flames, causing interference.\newline Compared to   each other, phase overlaps can occur, causing the flames to sometimes   strengthen and sometimes weaken. Depending on the phase difference, one flame   may be larger or smaller than the other. Since the phase difference changes   due to the fixed tone frequency, the ratio of the two flames also changes   constantly.&
Initial response from ChatGPT. To get a more concise answer, ChatGPT was asked to summarize the answer in one paragraph which lead to response R2.6\\

R2.6 & Die beiden brennenden Kerzen reagieren auf den kontinuierlichen Ton   des Lautsprechers durch Interferenz. Dabei verändern sich die Flammengrößen,   da die Schallwellen auf sie einwirken und unterschiedliche Phasenlagen   aufgrund der unterschiedlichen Entfernungen zum Lautsprecher erzeugen. Dies   führt dazu, dass sich die Flammen mal verstärken und mal abschwächen, da sich   Phasenüberlagerungen ergeben.                                                                                                                                                                                                                                                                                       & The two burning candles react to   the continuous sound of the loudspeaker by interference. In the process, the   flame sizes change as the sound waves act on them and create different phase   positions due to the different distances from the loudspeaker. This results   in the flames sometimes intensifying and sometimes weakening due to phase   overlaps.                                                                                                                                                                                                                                                                                                              & Overall best rated ChatGPT response (especially   regarding the linguistic quality) in the pilots and therefore chosen for the   main survey. \\\hline\hline
Q3       & Auf dem Boden eines mit Wasser gefüllten Glases liegen   Papierschnipsel. Sie rühren das Wasser mit einem Löffel um. Nach Herausnehmen   des Löffels beobachten Sie, dass die Papierschnipsel aufsteigen und sich nach   einiger Zeit in der Mitte des Bodens sammeln. Erklären Sie dieses Phänomen!                                                                                                                                                                                                                                                                                                                                                                                                            & Paper scraps are lying on the   bottom of a glass filled with water. You stir the water with a spoon. After   removing the spoon, you observe that the paper scraps rise and collect in the   middle of the bottom after some time. Explain this phenomenon!                                                                                                                                                                                                                                                                                                                                                                                                                      & High level of difficulty (transfer of   knowledge to a for the students unknown problem)             \\\hline
R3.1 & Dieses Phänomen wird durch die Strömungen im Wasser verursacht. Das   Umrühren des Wassers erzeugt Strömungen und Turbulenzen, die dazu führen,   dass die Papierschnipsel aufsteigen. Nach einer Weile sinken die   Papierschnipsel jedoch wieder auf den Boden des Glases, weil die Strömungen   nachlassen und das Papier durch sein eigenes Gewicht wieder nach unten sinkt.   In der Mitte des Bodens sammeln sie sich aufgrund der niedrigeren   Strömungsgeschwindigkeit.                                                                                                                                                                                                                                & This phenomenon is caused by the   currents in the water. Stirring the water creates currents and turbulence   that cause the paper scraps to rise. After a while, however, the paper scraps   sink back to the bottom of the glass because the currents subside and the   paper sinks back down due to its own weight. They collect in the middle of   the bottom because of the lower flow velocity.                                                                                                                                                                                                                                                                            & Excluded based on the pilot   results because similarly to response 3.4 (only slightly worse).  \\
R3.2 & Dieses Phänomen wird durch die Archimedesche Prinzip erklärt. Es   besagt, dass ein Körper in einer Flüssigkeit eine Auftriebskraft erfährt, die   gleich dem Gewicht des von ihm verdrängten Flüssigkeitsvolumens ist. Da die   Papierschnipsel leichter sind als das Wasser, das sie verdrängen, erfahren   sie eine Auftriebskraft und steigen auf. Die Papierschnipsel sammeln sich   schließlich in der Mitte des Bodens, weil sie ansonsten den Rand des Glases   berühren und somit den Auftrieb verringern würden.                                                                                                                                                                                      & This phenomenon is explained by   Archimedes' principle. It states that a body in a liquid experiences a   buoyant force equal to the weight of the volume of liquid it displaces. Since   the paper shreds are lighter than the water they displace, they experience a   buoyant force and rise. The paper shreds eventually collect in the center of   the bottom because they would otherwise touch the edge of the glass, thus reducing   buoyancy.                                                                                                                                                                                                                           & Excluded based on the pilot   results because similarly rated in the pilots as repsonse 3.3 but with a   similar explanation as the other responses. \\
R3.3 & Dieses Phänomen kann durch den Prozess des Sinkens und Steigens von   Luftblasen im Wasser erklärt werden. Beim Umrühren des Wassers mit dem Löffel   werden Luftblasen in das Wasser eingeführt, die sich auf den Papierschnipseln   sammeln und diese nach oben treiben lassen. Nach einiger Zeit sinken die   Papierschnipsel wieder, da die Luftblasen aus den Papierschnipseln entweichen   und die Papierschnipsel wieder ihr eigenes spezifisches Gewicht haben, das   sie zum Boden zurücktreibt. Daher sammeln sich die Papierschnipsel in der   Mitte des Bodens, weil die Luftblasen hier nicht mehr vorhanden sind und die   Papierschnipsel ihr eigenes spezifisches Gewicht haben.                & This phenomenon can be explained   by the process of sinking and rising of air bubbles in the water. When the   water is stirred with a spoon, air bubbles are stirred into the water, which   collect on the paper scraps and cause them to float upwards. After some time,   the paper scraps sink again as the air bubbles escape from the paper scraps   and the paper scraps have their own specific gravity again, which drives them   back to the bottom. Therefore, the paper scraps collect in the middle of the   bottom because the air bubbles are no longer present here and the paper   scraps have their own specific weight.                                      & ChatGPT response rated with medium   quality in the pilots and with a completely different explanation (air   bubbles) as all other responses, thus chosen for the main survey.   \\
R3.4 & Dieses Phänomen wird durch den sogenannten Bouyancy-Effekt erklärt.   Wenn Sie das Wasser mit dem Löffel umrühren, entsteht eine Strömung, die die   Papierschnipsel nach oben drückt. Die Papierschnipsel sinken jedoch wieder   ab, da ihr spezifisches Gewicht größer ist als das des Wassers und sie daher   wieder nach unten sinken. Die Papierschnipsel sammeln sich schließlich in der   Mitte des Bodens, da sie dort durch den natürlichen Ausgleich der Strömungen   zur Ruhe kommen.                                                                                                                                                                                                                & This phenomenon is explained by   the so-called Bouyancy effect. When you stir the water with the spoon, a   current is created that pushes the paper shreds upwards. However, the paper   scraps sink again because their specific weight is greater than that of the   water and they therefore sink back down. The paper scraps finally collect in   the middle of the bottom, as they come to rest there due to the natural   balancing of the currents.                                                                                                                                                                                                                      & Best rated ChatGPT response in the   pilots and therefore chosen for the main survey.  \\
R3.5 & Dieses Phänomen tritt auf, weil das Papier anfangs eine höhere Dichte   hat als das umgebende Wasser, wodurch es sinkt. Wenn es jedoch Wasser   aufnimmt und somit seine Dichte reduziert, steigt es auf. Schließlich wird es   jedoch wieder schwerer und sinkt aufgrund seiner größeren Dichte auf den   Boden, wo es sich in der Mitte sammelt.                                                                                                                                                                                                                                                                                                                                                              & This phenomenon occurs because the   paper initially has a higher density than the surrounding water, which causes   it to sink. However, when it absorbs water and thus reduces its density, it   rises. Eventually, however, it becomes heavier again and, due to its greater   density, sinks to the bottom, where it collects in the middle.                                                                                                                                                                                                                                                                                                                                  & Worst rated ChatGPT response in the   pilots and therefore chosen for the main survey. \\
R3.6& Der Löffel versetzt das Wasser in Rotation, sodass auf das Wasser   eine Zentrifugalkraft wirkt. Dieser Kraft ist durch die Reibung des Wassers   mit dem Glasboden eine Reibungskraft entgegengesetzt. Durch den hierdurch   entstehenden Geschwindigkeitsgradienten entsteht eine der Rotationsbewegung   überlagerte Sekundärströmung. Diese Strömung läuft von der Mitte des   Glasbodens nach oben, in höheren Schichten in Richtung Rand, am Rand des   Glases nach unten und am Glasboden zur Mitte. Die Papierschnipsel werden von   dieser Strömung mitgenommen. Nach Herausnehmen des Löffels lässt die Strömung   reibungsbedingt nach und die Schnipsel sammeln sich letztlich in der   Bodenmitte.& The spoon causes the water to   rotate, so that a centrifugal force acts on the water. This force is   countered by a frictional force due to the friction of the water with the bottom   of the glass. The resulting velocity gradient creates a secondary flow   superimposed on the rotational motion. This flow runs from the center of the bottom   of the glass upwards, in higher layers towards the edge, at the edge of the   glass downwards, and at the glass bottom towards the center. The paper scraps   are carried along by this flow. After removing the spoon, the flow decreases   due to friction and the shreds finally collect in the center of the bottom. & Sample solution; One expert reached out after the   main study had already begun and pointed out an imprecision in this sample solution. We could not implement that into the sample solution given to the   students anymore: The secondary flow is caused by an   unbalanced pressure gradient, which in turn is caused by a different centripetal force at the top and bottom (since the rotational speed is lower   due to friction at the bottom). A centrifugal force, on the other hand, is only effective when described in the comoving frame.\\
\hline\hline
\label{tab:qandr}
\end{longtable*}

\clearpage

\newpage

\bibliography{ChatGPT}

\end{document}